\documentclass[11pt,twoside]{article}

\usepackage{asp2004}
\usepackage{epsf}
\usepackage{psfig}
\usepackage{lscape}

\begin{document}
\title{Astrometric Monitoring of Stellar Orbits at the \\ Galactic Center with a Next Generation Large Telescope}   %%% Fill in title
\author{Nevin N.~Weinberg and Milo\v s Milosavljevi\'c} 
\affil{Theoretical Astrophysics, California Institute of Technology, 
Pasadena, CA 91125}
\author{Andrea M.~Ghez} 
\affil{Department of Physics and Astronomy, 
University of California, Los Angeles, CA 90095-1562}    %%% Fill in author affiliations

\begin{abstract} %%% Abstract to run on from here.
We show that with a Next Generation Large Telescope one can detect the accelerated motions of $\sim 100$ stars orbiting the massive black hole at the Galactic center. The positions and velocities of these stars will be measured to astrometric and spectroscopic precision several times better than currently attainable enabling detailed measurements of the gravitational potential in the neighborhood of the massive black hole. We show that the monitoring of stellar motions with such a telescopes enables: (1)  a measurement of the Galactic center distance $R_0$ to better than 0.1\% accuracy, (2) a measurement of the extended matter distribution near the black hole, including that of the exotic dark matter, (3) a detection of general relativistic effects due to the black hole including the prograde precession of stars and possibly the black hole spin, and (4) a detection of gravitational encounters between monitored stars and stellar remnants that accumulate near the Galactic center. Such encounters probe the mass function of the remnants.
\end{abstract}

%%% MAIN BODY OF TEXT GOES HERE. CONSULT "INSTRUCTIONS FOR AUTHORS USING
%%% LATEX2E MARKUP", SECTIONS 2.3-2.6 FOR HELP WITH EQUATIONS, FIGURES,
%%% AND TABLES.

%\section{}   %%% Top level section head (remove "%" symbol)
%\subsection{}   %%% Second level section head (remove "%" symbol)
%\subsubsection{}   %%% Lowest level section head (remove "%" symbol)
%\section*{}	%%% Unnumbered top level section head (remove "%" symbol)
%\subsection*{}   %%% Unnumbered second level section head (remove "%" symbol)

\section{Introduction}
Near-infrared monitoring of stellar sources in the inner $1\arcsec$ of the Galaxy with speckle imaging and adaptive optics techniques has enabled complete orbital reconstruction of several stellar sources orbiting the $\sim 4 \times 10^6 M_\odot$ black hole at the Galactic center \citep{Schoedel:02, Schoedel:03, Ghez:05}. Sources have been monitored with astrometric errors of a few milli-arcseconds, and radial velocity errors $< 50 \textrm{ km s}^{-1}$  (Eisenhauer et al. 2003; Ghez et al. 2003a), allowing the detection of the accelerated proper motions of $\sim 10$ stars. The star S0-2 has the shortest orbital period of $\sim15 \textrm{ yr}$. With current orbital data, the distance to the Galactic center $R_0$ is measured to an accuracy of $\approx5\%$ (\citealt{Eisenhauer:03, Eisenhauer:05}; see \citealt{Salim:99} for a discussion of the physical principles underlying the measurement of $R_0$ from the orbital data).

Although the stellar orbits have provided unequivocal proof of the existence of a massive black hole at the Galactic center, the matter content in the vicinity of the black hole remains largely unknown.  Due to the short radial diffusion time as compared with the age of the bulge, a large number of massive compact remnants ($5-10M_\odot$ black holes) could have segregated into the dynamical sphere of influence of the black hole \citep{Morris:93,Miralda:00}.  Furthermore, adiabatic growth of the massive black hole could have compressed a pre-existing distribution of cold dark matter (CDM) \citep{Ipser:87,Quinlan:95,Gondolo:99} and stars \citep{Peebles:72,Young:80} into a density ``spike'', although a variety of dynamical processes are capable of destroying such a spike \citep{Ullio:01,Merritt:02,Gnedin:04,Merritt:04}. If the CDM consists of weakly-interacting massive particles (WIMPs), a sustained CDM spike might produce detectable WIMP annihilation radiation.

We show here that the finer angular resolution and increased light-gathering power of a Next Generation Large Telescope (NGLT) will enable the detection of accelerated proper motions of $\sim 100$ stars with astrometric and spectroscopic errors several times smaller than currently possible. By fitting to mock NGLT orbital data we determine the constraints such orbits place on the gravitational potential near the massive black hole at the Galactic center (see also \citealt{Weinberg:05}). We demonstrate that the observations will yield a measurement of the density profile in a dark matter spike. They will also reveal the lowest order general relativistic effects and possibly higher order effects such as the black hole spin. We show that $R_0$ will be measured to better than $0.1\%$ precision, helping to place tight constraints on the shape of the Galactic dark matter halo. We also calculate the rate at which perturbations from massive remnants that accumulate near the Galactic center deflect the observed stellar orbital motions and describe how the remnants' mass function can thereby be extracted from the monitoring data. We find that if stellar-mass black holes dominate the matter density then in 10 yr of monitoring at an astrometric resolution of $0.5 \textrm{ mas}$ approximately 10\% of all monitored stars will experience detectable deflections in their orbital motions. 

\section{Monitoring Stars near the Galactic Center with an NGLT}
\subsection{Astrometric and Spectroscopic Limit}
\label{sec:astrometry}

The ability to extract information about the Galactic center environment via the monitoring of stellar motions is limited by the astrometric and spectroscopic precision of the measurements. We now estimate each in turn.

Adaptive optics will enable an NGLT to operate near its diffraction limit. The astrometric limit achievable with adaptive optics on an NGLT scales with telescope aperture as $D^{2/3}$ \citep{Shao:92}. Since the astrometric limit of bright stellar sources at the Galactic center achieved today with 10 meter telescopes is $\delta \theta_{10} \sim 1 \textrm{ mas}$, a 30 meter NGLT should achieve $\delta \theta_{30} \sim 0.5 \textrm{ mas}$. This is a conservative estimate as future adaptive optic platforms may provide astrometric limits $\sim 1\%$ of the diffraction limit (M. Fitzgerald, private communication), corresponding to $\delta \theta_{30} \sim 0.1 \textrm{ mas}$ in the $K$-band. 

Integral field spectroscopy with an NGLT is expected to provide a spectral resolution $R = \lambda / \Delta \lambda \sim 1- 2 \times 10^4$. Thus, with an NGLT one can measure stellar radial velocities to an accuracy of $\delta v \sim 10 \textrm{ km s}^{-1}$. This too is a conservative estimate. The faint, cool stars detectable with an NGLT will likely exhibit rich spectral features including molecular lines, enabling high spectral resolution studies. 
For example, \citet{Figer:03} measured the radial velocities of 85 cool stars in the central parsec of the Galaxy with velocity errors of $\sim 1 \textrm{ km s}^{-1}$. Thus, an NGLT may achieve $\delta v$ ten times smaller than our above estimate.

\subsection{Confusion Limit}

\begin{figure}[!ht]
\plotfiddle{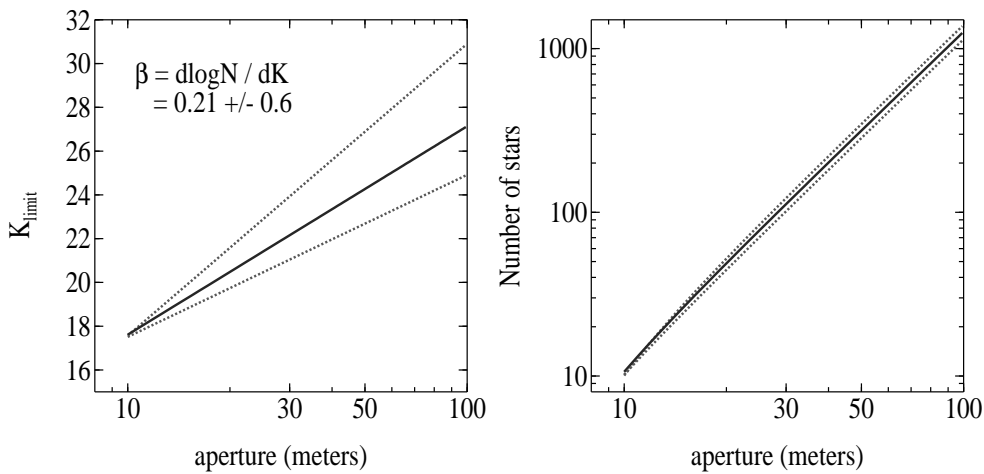}{2.3in}{270}{55}{55}{-220}{250}
\caption{ The $K$-band magnitude limit and number $N$ of stars with detectable orbital motions as a function of the aperture of a diffraction limited NGLT. Results are shown for power-law $K$-band luminosity functions normalized to observations by \citet{Schoedel:03} with slopes matching the $\sim 2 \sigma$ range found by \citet{Genzel:03a}. After \citet{Weinberg:05}.
\label{fig:number}}
\end{figure}

The greatest obstacle to detecting and following hitherto unseen stars in the central $1\arcsec$ of the Galaxy is stellar crowding.  The light of bright stars and underlying faint stars flood the pixel elements and impose a limit to the faintest detectable star. The limit depends on the telescope specifications and on the luminosity function of the stars in the field. In \citet{Weinberg:05} we estimate the minimum luminosity a star must have in order to be detected and monitored with an NGLT by evaluating the confusion noise as a function of the stellar $K$-band luminosity function (KLF). We normalize the KLF to the observations of \citet{Schoedel:03} of bright stars within the inner arcsecond and use the best-fit KLF slope from \citet{Genzel:03a}. We integrate the KLF over stars brighter than the derived minimum luminosity to obtain an estimate of the total number of stars that can be detected and monitored with an NGLT. We find that a 30 meter NGLT will resolve $N \sim 100$ stars out to a magnitude limit of $K \sim 22$ (Figure \ref{fig:number}). For such an NGLT, confusion noise excludes the detection of stars with semi-major axes smaller than $a \sim 300 \textrm{ AU}$ and the acceleration threshold below which orbital motion is not detectable excludes stars with  $a > 3000 \textrm{ AU}$.

\section{Modeling the Orbital Dynamics}

Our goal is to estimate the uncertainties in the parameters that describe the gravitational potential through which the monitored stars move. We do so by first generating mock orbital data assuming a particular model for the potential (e.g., black hole mass, extended matter density profile, etc.) and a particular observing program (e.g., time baseline of 10 yr, 10 observations per year per star with measurement errors $[\delta \theta, \delta v] = [0.5 \textrm{ mas}, 10 \textrm{ km s}^{-1}]$). The orbital parameters are drawn from the distribution function of a realistic $r^{-3/2}$ power-law density profile assuming randomly oriented orbits and considering only those orbits detectable with an NGLT (e.g., $N = 100$ stars with semi-major axes in the range $300 \textrm{ AU} < a < 3000 \textrm{ AU}$). 

\subsection{Equations of Motion}
An orbit fitting routine must model various non-Keplerian motions in order to extract information about the gravitational potential near the massive black hole. Effects that give rise to deviations from Keplerian motions include the Newtonian retrograde precession of an orbit due to the presence of an extended matter distribution, the relativistic prograde precession, and the frame dragging effects due to the black hole spin.  The post-Newtonian approximation to the geodesic equation for test particles orbiting a spherically symmetric mass accounts for the above effects and provides an appropriate description of the observable motions (see \citealt{Weinberg:05}). 

We also examine the prospect of directly measuring the mass function of the remnants by looking for changes in orbital semi-major axis due to the gravitational interaction between monitored stars and background massive remnants. Such interactions will be of two types: strong encounters and weak encounters. The former correspond to cases where the impact parameter of an encounter satisfies $d < Gm_\star/v_{\rm rel}^2$, where $m_\star$ is the mass of the background star and $v_{\rm rel}$ is the relative velocity of the encounter, causing a large change in semi-major axis. We calculate the rate of strong encounters by extending the analysis of \citet{Goodman:83}, who considered the encounter rate in globular clusters. Weak encounters describe the slow diffusion of a star's orbital energy due to the continual small angle scattering by the background stellar distribution, whose total potential is the sum of discrete point-mass potentials. We solve the orbit-averaged Fokker-Planck equation \citep{Cohn:78} to calculate the rate at which weak encounters change a star's semi-major axis. The rate of strong and weak encounters is proportional to the mass of the remnants.

\subsection{Parameter Estimation}

A star's projected orbit is described by six phase space parameters. The potential contributes six additional parameters: the black hole mass, its 3-dimensional position, and the normalization and slope of the extended matter distribution. The dimension of the parameter space of our model is therefore $J = 6N + 6$. Thus, to extract parameter constraints from the $N \sim 100$ stars detectable with a 30 meter NGLT requires exploring a $J \sim 600$ dimensional parameter space. Since the computational cost of the grid-based approach increases exponentially with $J$,  parameter estimation on a $J$-dimensional grid is not practical for $N \ga 3$. Instead we use the Markov Chain Monte Carlo (MCMC) method which provides an efficient means of mapping out the probability distribution of large dimensional models. The details of our implementation of the MCMC method is given in \citet{Weinberg:05}.

\section{Results}

We show how well an NGLT constrains the mass and distance to the massive black at the Galactic center and the extended matter distribution of stars and dark matter around the black hole. We also demonstrate that relativistic effects are detectable with an NGLT. Finally, we show that interstellar interactions between monitored stars and background massive remnants will be observed with an NGLT if the population of stellar-mass black holes predicted by theory exists.  Although we show results for a diffraction limited 30 meter NGLT with  $(\delta \theta, \delta v) = (0.5 \textrm{ mas}, 10 \textrm{ km s}^{-1})$ and $(\delta \theta, \delta v) = (0.1 \textrm{ mas}, 1 \textrm{ km s}^{-1})$, since the parameter uncertainties scale with measurement error $\sigma$ (i.e., $\delta \theta$ and $\delta v$) and number of monitored stars $N$ as $\sigma / N^{1/2}$, the results can be used to describe the capabilities of an NGLT with different specifications. For example, based on Figure \ref{fig:number} a 100 meter NGLT will detect ten times as many stars as a 30 meter NGLT. Thus, if a 100 meter telescope has astrometric and spectroscopic errors that are smaller than those of a 30 meter telescope by a factor of five, the parameter uncertainties will be $\sim 10\times$ smaller.

\subsection{Measuring $M_{\rm bh}$ and $R_0$}
\begin{figure}[!ht]
\plotfiddle{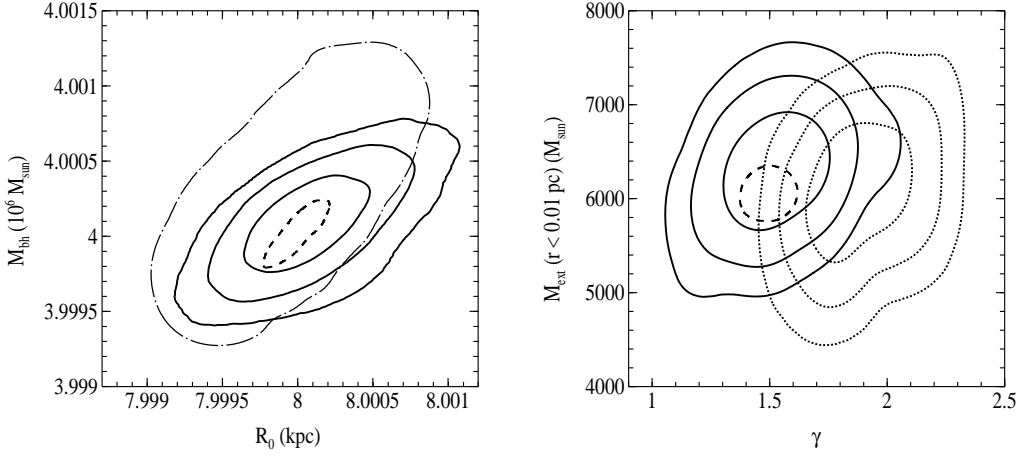}{2.3in}{270}{50}{50}{-210}{240}
\caption{Left panel: The constraints on $M_{\rm bh}$ and $R_0$ obtainable with an NGLT. The solid contours show the 68\%, 95\%, and 99.7\% confidence levels assuming an astrometric limit of $\delta \theta = 0.5 \textrm{ mas}$ and a spectroscopic limit of $\delta v = 10 \textrm{ km s}^{-1}$ for a mock sample of stars detectable with an NGLT. The line-dot contour shows the 99.7\% confidence level for a different draw of mock orbits. The dashed contour shows the 99.7\% confidence level for smaller astrometric and spectroscopic limits of $\delta \theta = 0.1 \textrm{ mas}$ and $\delta v = 2 \textrm{ km s}^{-1}$.  Right panel: The constraint on the extended matter distribution obtainable with an NGLT.  The input models have power-law slope of $\gamma = 1.5$ and $\gamma = 2$ and an input enclosed mass of $6000 M_\odot$ within $0.01\textrm{ pc}$. The line styles are the same as in the left panel. After \citet{Weinberg:05}. \label{fig:ngltcontour}}
\end{figure}

For an astrometric limit of $\delta \theta =0.5 \textrm{ mas}$ and a spectroscopic limit of $\delta v = 10 \textrm{ km s}^{-1}$ the fractional uncertainties in $M_{\rm bh}$ and $R_0$ are less than 0.1\% at the 99.7\% level (Figure \ref{fig:ngltcontour}), a factor of $\sim 100$ times better than present uncertainties. For astrometric and spectroscopic limits that are a factor of five smaller the fractional uncertainties in $M_{\rm bh}$ and $R_0$ are smaller by almost a factor of five. 

As shown by \citet{Olling:00, Olling:01} measuring the Galactic constants $R_0$ and the Galactic rotation speed $\Theta_0$ to high accuracy will constrain the shortest-to-longest axis ratio $q = c /a$ of the Galactic dark matter halo to similar accuracy. The shape parameter $q$ is an important diagnostic of dark matter models and structure formation and is currently poorly constrained in all galaxies including the Milky Way.

\subsection{Measuring the Extended Matter Distribution}
\label{sec:resultextended}
In Figure \ref{fig:ngltcontour} we show the constraints an NGLT will place on the extended matter distribution for input power-law models with $M_{\rm ext} (r < 0.01 \textrm{ pc}) = 6000 M_\odot$ and $\gamma = 1.5$ or $\gamma = 2$. We find that an NGLT will measure $M_{\rm ext}$ and $\gamma$ to $20 - 30\%$ accuracy. We find that an extended matter distribution is detectable (i.e., observations yield a lower bound) for $\delta \theta = 0.5 \textrm{ mas}$ and $\delta v = 10 \textrm{ km s}^{-1}$ as long as $M_{\rm ext} (r < 0.01 \textrm{ pc}) \ga 1500 M_\odot$.  Thus, if the dark matter distribution matches theoretical expectations and forms a density spike, its influence on the orbits will be detectable with an NGLT. A detection would constitute a measurement of the gravitational influence of dark matter on the smallest scales yet.

\subsection{Measuring Relativistic Effects}
\label{sec:resultsrelativ}
 
The relativistic prograde precession (RPP) causes a pericenter advance per revolution of  $\Delta \phi_{\rm pro} = 3 \pi R_s/a (1 - e^2)$, where $R_s=2GM_{\rm bh}/c^2$ is the Schwarzschild radius of the black hole (see \citealt{Weinberg:72}). The apparent apocenter shift per revolution caused by the RPP is $\Delta s \approx \Delta \phi_{\rm pro} a (1+e) / R_0 = 3 \pi R_s/ R_0 (1 - e)$, which corresponds to an apparent shift of $\sim 1 \textrm{ mas}$ for the star S0-2. Consider an orbit seen face on and observed for $N_{\rm orb}$ complete periods. Since the precession angles per revolution add linearly, the signal-to-noise from the RPP is $\emph{S}_{\rm pro} \sim\Delta s N_{\rm orb}/\delta \theta$, or
\begin{equation}
\emph{S}_{\rm pro} \sim  0.1 \frac{N_{\rm orb}}{1-e} 
\left( \frac{M_{\rm bh}}{4 \times 10^6 \; M_\odot}\right)
                              \left(\frac{R_0}{8 \; \rm{kpc}} \right)^{-1}
                              \left(\frac{\delta \theta}{1 \; \rm{mas}} \right)^{-1}. 
\end{equation}
If only one star with $e > 0.96$ is monitored over a single period at an astrometric accuracy $\delta \theta = 0.5 \textrm{ mas}$, the RPP shift will be measured to 5-$\sigma$ accuracy. Several such stars are expected to be detected in a 30 meter NGLT's sample of 100 stars.

Frame dragging effects due to a spinning black hole also cause a precession of a stellar orbit's pericenter. The spin precession per revolution for a star orbiting a black hole with spin angular momentum $J$ is given approximately by (see \citealt{Weinberg:72}, equation (9.5.22); note different notation)
\begin{equation}
\Delta \phi_{\rm spin} \approx -8 \pi j \left(\frac{G M_{\rm bh}}{c L}\right)^3 \cos \psi 
\end{equation}
where $\psi$ is the angle between the orbital angular momentum vector and the black hole spin axis and $0 \leq j \equiv cJ / GM_{\rm bh}^2 \leq 1$ is the black hole spin parameter. The black hole spin induces an apocenter shift that is smaller than the RPP shift by a factor of $\sim v/c$. For an orbit observed face-on the signal-to-noise from a spin-induced apocenter shift is
\begin{eqnarray}
\label{eq:SNspin}
\emph{S}_{\rm spin}  & \approx & 0.001 \frac{j N_{\rm orb} \cos \psi}{\sqrt{(1+e)(1-e)^3}} \left( \frac{M_{\rm bh}}{4 \times 10^6 \; M_\odot}\right)^{3/2}  
		\nonumber \\ & & \times
                	          \left( \frac{a}{1000 \rm{ AU}}\right)^{-1/2}
                              \left(\frac{R_0}{8 \; \rm{kpc}} \right)^{-1}
                              \left(\frac{\delta \theta}{1 \; \rm{mas}} \right)^{-1}. 
\end{eqnarray}
In order to be detectable a spin-induced orbital precession requires an astrometric
precession $\delta \theta \la 0.05 \textrm{ mas}$ or the favorable detection of a star on a highly compact and eccentric orbit (see also \citealt{Jaroszynski:98,Fragile:00}).

To confirm the above order of magnitude estimates we generated mock orbits using the post-Newtonian corrections to the equations of motion and fit to these orbits using the MCMC method while treating the speed of light as a parameter in the model. We found that the RPP effect is measurable for $\delta \theta = 0.5 \textrm{ mas}$ (speed of light constrained to 5\% accuracy) while the spin induced precession is not.
\begin{figure}[!h]
\plotfiddle{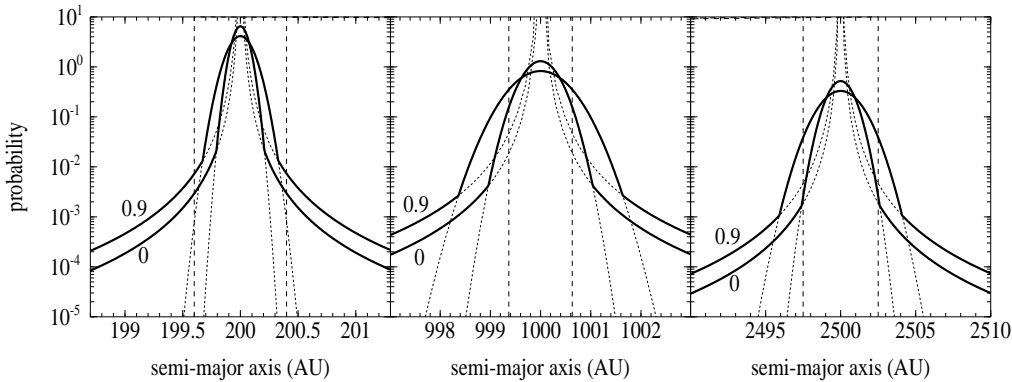}{1.8in}{270}{57}{57}{-230}{230}
\caption{ The probability that after $T = 10 \textrm{ yr}$ the semi-major axis of a star is changed from an initial value $a_i$ to a final value $a_f$ due to strong and weak encounters. The three panels show results for the cases $a_i = 200$, 1000, $2500 \textrm{ AU}$ with $e = 0$ and $e = 0.9$. The dotted lines show the separate contributions from strong and weak encounters and the solid lines show the net probability. The vertical dashed lines are the uncertainties in $a_i$ assuming $\delta \theta = 0.5 \textrm{ mas}$.\label{fig:prob}}
\end{figure}

\subsection{Interstellar Interactions}

\citet{Milosavljevic:06} estimate the rate at which monitored stars experience detectable encounters with background stars and stellar remnants (see also \citealt{Weinberg:05}).  Such encounters will manifest themselves as changes in a star's orbital energy, i.e., semi-major axis (Figure \ref{fig:prob}), at a rate proportional to the mass of the background sources. We found that if the background sources are dominated by stellar-mass black holes, as predicted by estimates of mass segregation in the vicinity of a massive black hole \citep{Morris:93,Miralda:00}, approximately 30 encounters will be detected over a ten year baseline for $\delta \theta = 0.5 \textrm{ mas}$. 

\section{Conclusions}

We have shown that the monitoring of stellar orbits around the massive black hole at the Galactic center at the high astrometric and spectroscopic resolution attainable with an NGLT enables one to probe the deep gravitational potential of the region. Many exciting measurements are achievable even for modest (factor of a few) improvements over the astrometric capabilities of current 10 meter class telescopes. 
The future success of Galactic center research greatly depends on advances in the astrometric capabilities of the next generation of telescopes.

%%% THE BIBLIOGRAPHY
%%%
%%% CONSULT SECTION 3 OF "INSTRUCTIONS FOR AUTHORS" FOR HOW TO USE NATBIB.
%%% AUTHORS ARE ENCOURAGED TO USE EITHER THE "THEBIBLIOGRAPY" ENVIRONMENT
%%% BY UNCOMMENTING (DELETING THE "%" SYMBOL) THE COMMANDS BELOW, OR BY
%%% USING THE BIBTEX ENVIRONMENT. TO FIND OUT WHICH IS APPLICABLE TO YOUR
%%% CONTRIBUTION, CONSULT THE VOLUME EDITORS FOR YOUR PROCEEDINGS.
%%%

\end{document}